\documentclass{article}
\usepackage{graphicx}
\usepackage{amsmath}
\usepackage[ruled]{algorithm2e}
\usepackage{fullpage}
\usepackage{authblk}
\usepackage{url}

\SetAlFnt{\small}
\SetAlCapFnt{\small}
\SetAlCapNameFnt{\small}
\SetAlCapHSkip{0pt}
\IncMargin{-\parindent}
\SetKwProg{Fn}{Function}{}{end}
\SetKwFunction{MORTONGENERAL}{MortonGeneral}
\SetKwFunction{HILBERTGENERAL}{HilbertGeneral}
\SetKwFunction{GENERATE3D}{Generate3D}

\title{Generalised 3D Morton and Hilbert Orderings}
\author{David Walker}

\affil{UTC Research Institute, University of Tennessee at Chattanooga, TN 37403}
\date{September 2023}

\begin{document}

\maketitle

\section{Introduction}
This document describes algorithms for generating general Morton and Hilbert orderings for three-dimensional data volumes.

\section{Morton Ordering}
It is well-known that the mapping between row-major and Morton orderings for a three-dimensional cube of size $N\times N\times N$, where $N=2^n$, can be expressed as an interleaving of the bits of the column, row, and slab indices.

\begin{figure}[ht]
  \centering
  \includegraphics[width=0.7\linewidth]{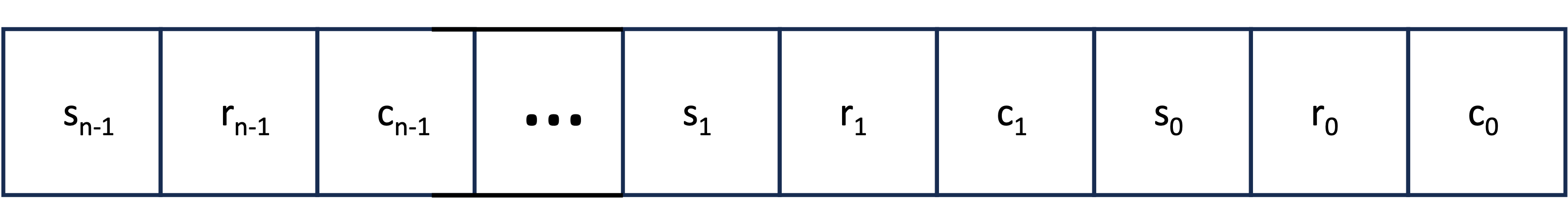}
  \caption{The bits of the Morton index  for a 3D cube of size $2^n$.}
  \label{fig:mortonbits}
\end{figure}

If the requirement of cubical data is relaxed, but the number of rows, columns, and slabs is still a power of 2, then the Morton ordering in this case is a modification of the cubical case. Suppose the size of the data is $2^p\times 2^n\times 2^m$, and assume without loss of generality that $m\le n\le p$. Then, the $3m$ least significant bits of the Morton ordering are the interleaved bits of the column, row, and slab indices. The next $2(n-m)$ bits of the Morton index are generated by interleaving bits $m$ to $n-m-1$ of the row and slab bits. Finally, the upper $p-n$ bits of the Morton index are the upper $p-n$ bits of the slab index.

\begin{figure}[ht]
  \centering
  \includegraphics[width=0.9\linewidth]{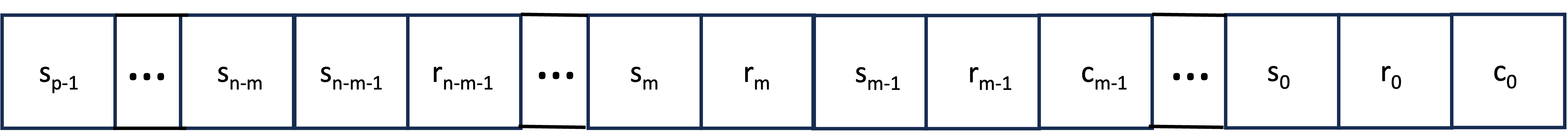}
  \caption{The bits of the Morton index for a 3D block of size $2^p\times 2^n\times 2^m$.}
  \label{fig:mortonnoncube}
\end{figure}

Now consider the general case in which the number of columns, rows, and slabs are $M$, $N$, and $P$, respectively. Now let $c$, $r$, and $s$ denote the following:
\[
c=\lfloor\log_2 M\rfloor,\qquad  r=\lfloor\log_2 N\rfloor,\qquad s=\lfloor\log_2 P\rfloor.
\]
and let $C=2^c$, $R=2^r$, and $S=2^s$. The 3D space of data indices is divided into octants of unequal size, with octant 0 being of size $S\times R\times C$ and being in the lower left front corner. This determines the size of the other seven octants. The mapping from row-major to Morton ordering is performed as in Fig.~\ref{fig:mortonnoncube}, and the other seven octants are then mapped recursively using the same process. The algorithm is outlined in Alg.~\ref{alg:one}. Each call to \MORTONGENERAL maps one octant (O$_0$) of size $S\times R\times C$ and adds it to the Morton path, {\it MPath}, starting at location ($S_0,R_0,C_0$).

\begin{figure}[pt]
  \centering
  \includegraphics[width=0.6\linewidth]{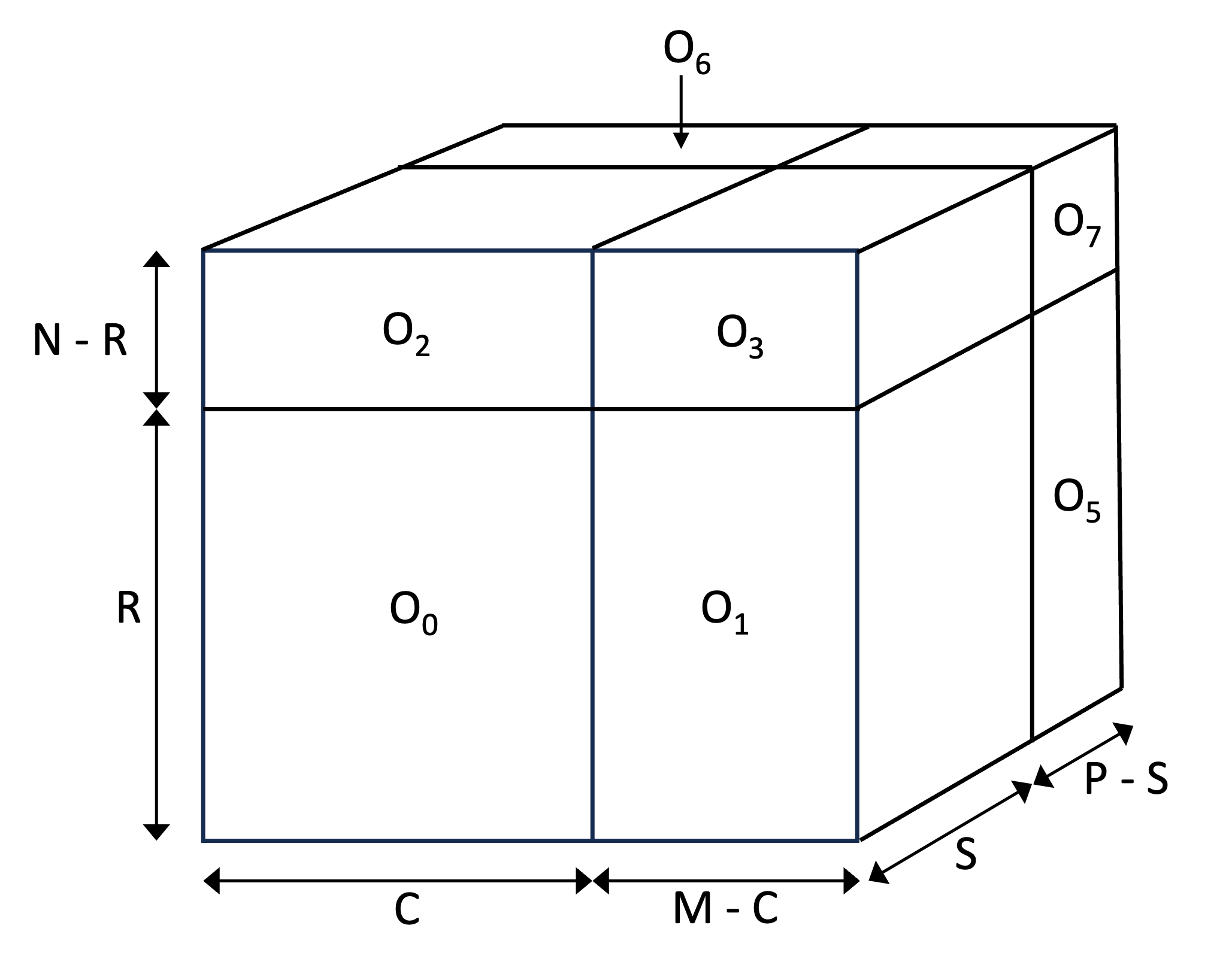}
  \caption{The division of a $P\times N\times M$ data array into octants, labelled O$_0$ to O$_7$. Octant 4 is occluded.}
  \label{fig:mortonnonblock}
\end{figure}

\begin{algorithm}[hpt]
\SetAlgoNoLine
\DontPrintSemicolon
\Fn{\MORTONGENERAL{$M,N,P,C0,R0,S0,Mpath$}}{
\KwIn{Size $(M,N,P)$ and location $(C0,R0,S0)$.}
\KwOut{Array of Morton indices, $MPath$.}
\If{($M*N*P=0$)}{
    return\;
}
Evaluate size $C, R, S$ of octant 0\;
Evaluate Morton indices for octant 0 starting at $(C0,R0,S0)$.\;
\MORTONGENERAL($M-C,R,S,C0+C,R0,S0,MPath$)\;
\MORTONGENERAL($C,N-R,S,C0,R0+R,S0,MPath$)\;
\MORTONGENERAL($M-C,N-R,S,C0+C,R0+R,S0,MPath$)\;
\MORTONGENERAL($C,R,P-S,C0,R0,S0+S,MPath$)\;
\MORTONGENERAL($M-C,R,P-S,C0+C,R0,S0+S,MPath$)\;
\MORTONGENERAL($C,N-R,P-S,C0,R0+R,S0+S,MPath$)\;
\MORTONGENERAL($M-C,N-R,P-S,C0+C,R0+R,S0+S,MPath$)\;
}
\caption{Algorithm for mapping a row-major ordering to a Morton ordering.}
\label{alg:one}
\end{algorithm}

\begin{figure}[ht]
  \centering
  \includegraphics[width=1\linewidth]{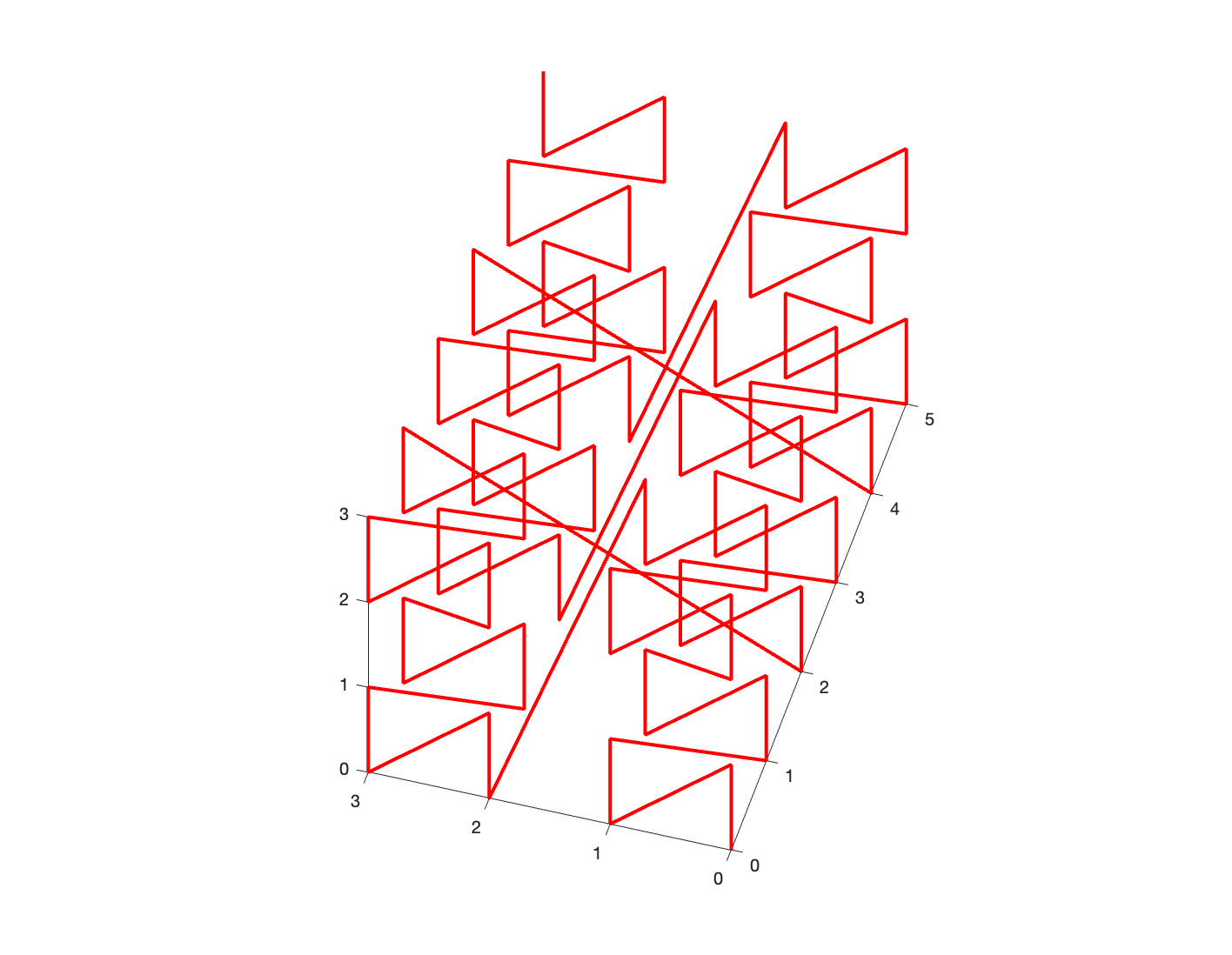}
  \caption{General Morton ordering for the case $P=6$, $N=4$, and $M=4$.}
  \label{fig:morton644}
\end{figure}

An example of a general Morton ordering for the case $6\times 4\times 4$ is shown in Fig.~\ref{fig:morton644}

\section{Hilbert Ordering}
This section is based on the code made available by Dr Jakub \v{C}erven\'{y}~\cite{jakub}. This approach requires the size of the data volume to be even in each dimension in order to avoid diagonal steps.  

\begin{figure}[hpt]
  \centering
  \includegraphics[width=0.8\linewidth]{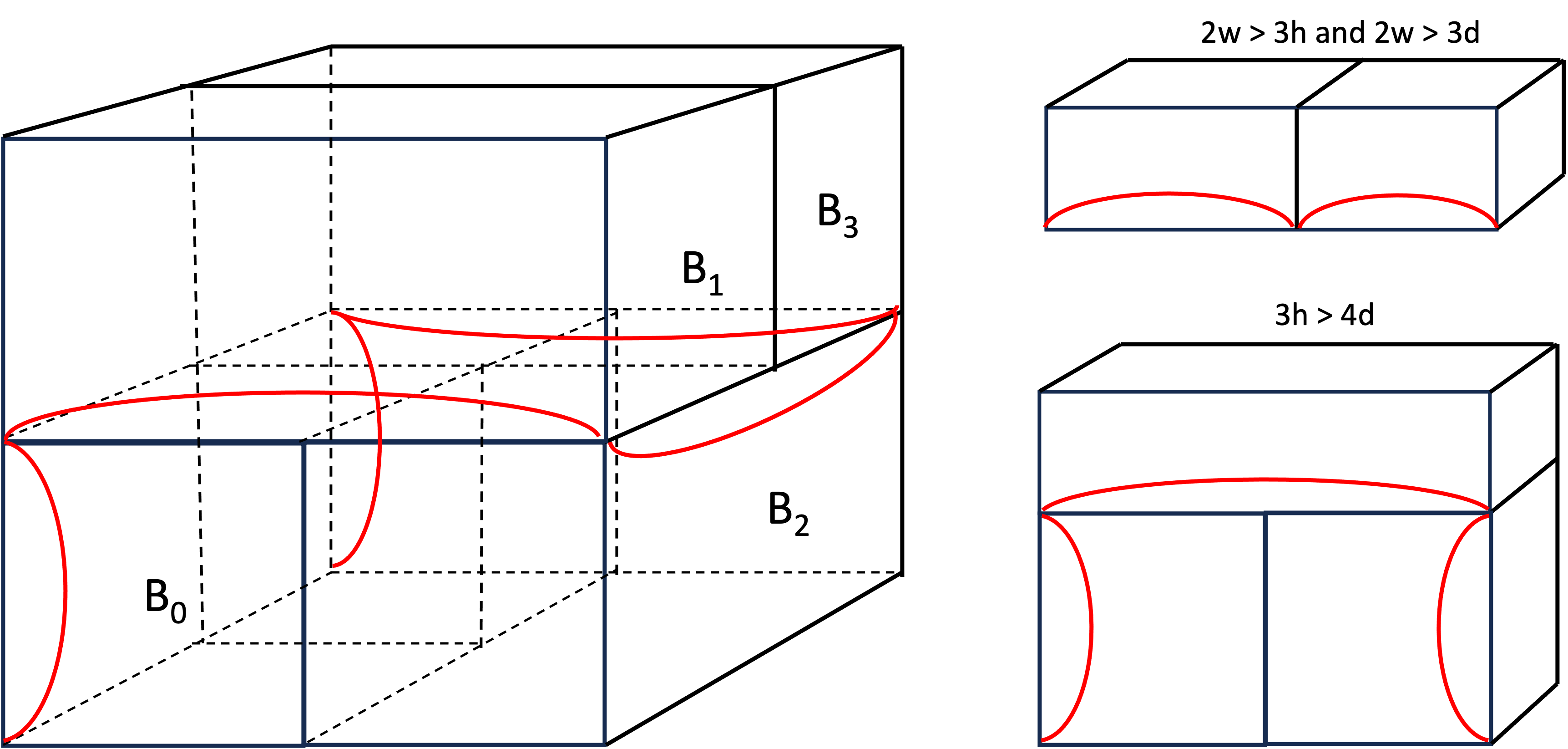}
  \caption{{\bf Left}: the division of a $P\times N\times M$ data array into five blocks, labelled B$_0$ to B$_4$, as described in the text. Block B$_4$ is the back, lower, left block. {\bf Top right}: the special case of a wide, thin block. {\bf Bottom right}: the special case of a block with height sufficiently greater than depth. The red curves represent the entry and exit points in each block of the Hilbert path.}
  \label{fig:GilbertBlocks}
\end{figure}

In general, a data volume of size $P\times N\times M$ is be divided into five blocks, except when one of the dimensions is sufficiently larger than the others (these special cases are considered below). Each block is then processed recursively. The recursion ends when at least two of the block's dimensions are of size 1, when $k$ points are added to the Hilbert path, where $k$ equals the longest dimension of the block. If the data volume is denoted by B[0:P-1,0:N-1,0:M-1],then the five blocks are illustrated in the left of Fig.~\ref{fig:GilbertBlocks}, and are as follows:
\setlength{\itemsep}{7pt}
\begin{itemize}
\setlength{\itemsep}{0pt}
    \item Block 0 is $B[0:d-1,0:h-1,0:w-1]$.
    \item Block 1 is $B[0:d-1,h:N-1,0:M-1]$.
    \item Block 2 is $B[0:P-1,0:h-1,w:M-1]$.
    \item Block 3 is $B[d:P-1,h:N-1,0:M-1]$.
    \item Block 4 is $B[d:P-1,0:h-1,0:w-1]$.
\end{itemize}

\begin{figure}[ht]
  \centering
  \includegraphics[width=0.8\linewidth]{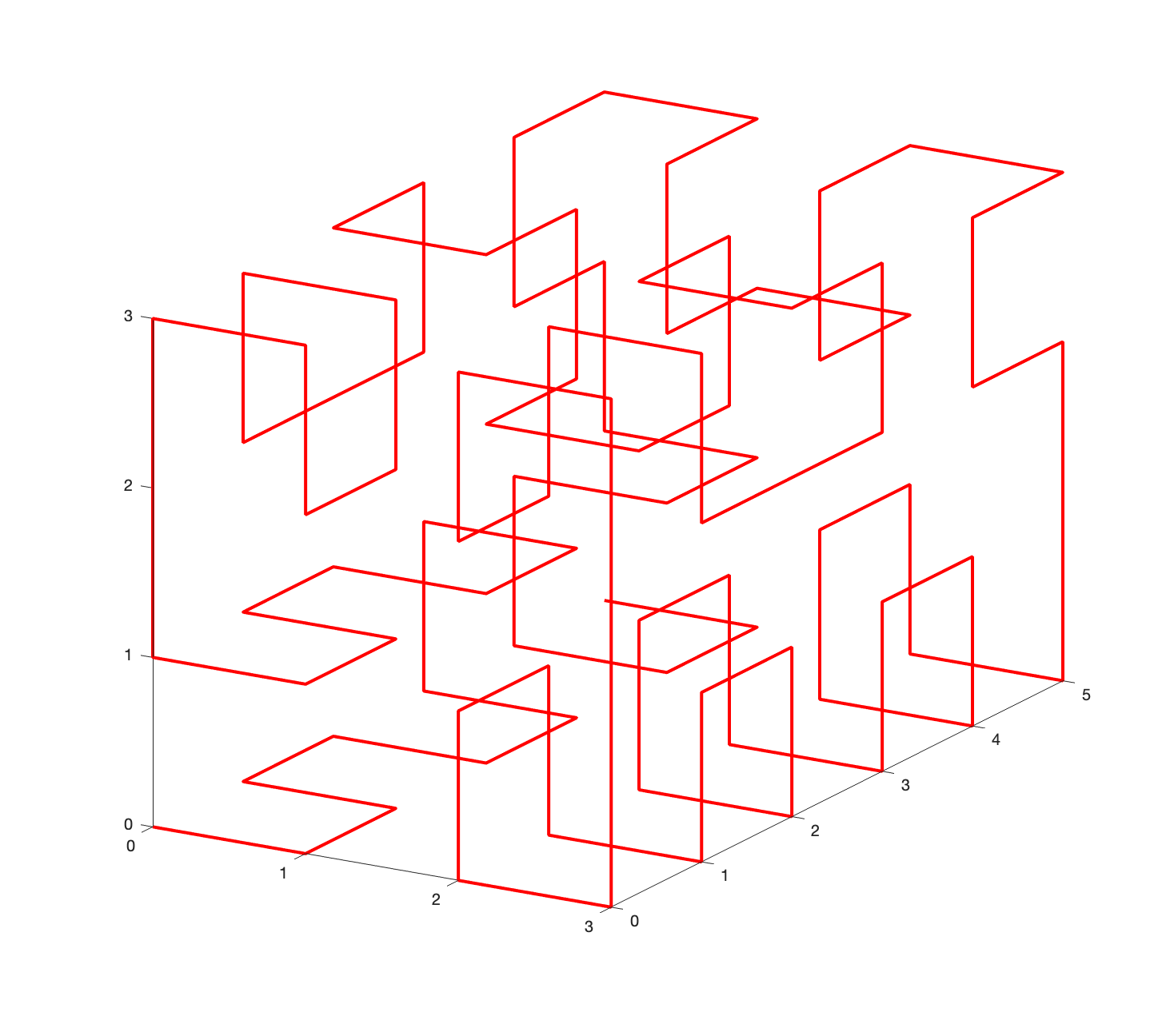}
  \caption{General Hilbert ordering for the case $P=6$, $N=4$, and $M=4$.}
  \label{fig:gilbert644}
\end{figure}

The value of $d$ is $\lfloor P/2\rfloor$; however, if this is an odd number it is adjusted by the addition or subtraction of 1. The values of $h$ and $w$ are defined and adjusted in a similar way. An example for the case $6\times 4\times 4$ is shown in Fig.~\ref{fig:gilbert644}. Block 0 is of size $4\times 2\times 2$; block 1 is $4\times 2 \times 4$; block 2 is $6\times 2\times 2$; block 3 is $2\times 2\times 4$; and block 4 is $2\times 2\times 2$.

If the current block is of size $d\times h\times w$, then three special cases must be considered. 
\begin{enumerate}
    \item If $2m>3h$ and $2m>3d$ then the current block is split in half orthogonal to the longest dimension, as shown in the top right of Fig.~\ref{fig:GilbertBlocks}.
    \item Otherwise, if $3h>4d$ then the current block is divided into three smaller blocks. The first division is made orthogonal to the height dimension. Then the lower of the resulting blocks is divided orthogonal to the width dimension, as shown in the bottom right of Fig.~\ref{fig:GilbertBlocks}.
    \item Otherwise, if $3d>4h$ then the current block is divided into three smaller blocks. The first division is made orthogonal to the depth dimension. Then the nearer of the resulting blocks is divided orthogonal to the width dimension. This is the same as special case 2, except the roles of height and depth are reversed.
\end{enumerate}

The Hilbert path starts at location $(0,0,0)$. Subsequent points are added to the path by moving unit distance in the current direction. The current direction is stored as the first column of the $3\times 3$ orientation matrix, with the other two columns being vectors orthogonal to the current direction. There are six possible current directions, corresponding to forward and backwards in the slab, row, and column dimensions. Thus, a point is added to the path by adding one of the six vectors $(\pm 1,0,0)$, $(0,\pm 1,0)$, and ($0,0,\pm 1)$ to the current position. The current direction can be deduced from the red lines shown in Fig.~\ref{fig:GilbertBlocks}. Thus, going from B$_0$ to B$_1$  (and from B$_2$ to B$_3$) updates the current position by adding $(0,1,0)$, whereas going from B$_1$ to B$_2$ (and from B$_3$ to B$_4$) adds $(0,-1,0)$ to the current position.

\section{Hybrid Orderings}
The data volume of size $P\times N\times M$ may be divided into blocks of size $p \times n\times m$, where $p$ is a divisor of $P$ $n$ is a divisor of $N$, and $m$ is a divisor of $M$. Different orderings may then be applied within the blocks, and between them. For example, a row-major ordering could be applied within the blocks, and then the blocks could have a Morton ordering. The size of the blocks is a parameter that can be varied to get the best cache performance.

\section{Concluding Remarks}
This short note has shown how to extend the Morton and Hilbert orderings for 3D data volumes of arbitrary size, although the Hilbert case requires an even number of data points in each dimension. Future work will investigate other approaches to the general Hilbert case, and the impact of the general Morton and Hilbert orderings on application performance. The use of space-filling curves in other geometries, such as spheres and pyramids, will also be considered.

\section*{Acknowledgements}
Funding in part is acknowledged from NSF Grant CCF-1918987, , as well as and the U.S. Department of Energy's National Nuclear Security Administration (NNSA) under the Predictive Science Academic Alliance Program (PSAAP-III), Award DE-NA0003966.

Any opinions, findings, and conclusions or recommendations expressed in this material are those of the authors and do not necessarily reflect the views of the 
National Science Foundation, the U.S.\hbox{} Department of Energy's National Nuclear Security Administration.

\bibliographystyle{plain}
\bibliography{main}

\begin{thebibliography}{1}

\bibitem{jakub}
Jakub Cerveny.
\newblock Generalized {H}ilbert space-filling curve for rectangular domains of arbitrary sizes.
\newblock \url{https://github.com/jakubcerveny/gilbert/}, 2021.
\newblock Accessed September 26, 2023.

\end{thebibliography}

\end{document}